Saint-Pierre en Gallicante

# Bibliothèque de la communauté assomptionniste

## Saisie informatique et classement Dewey

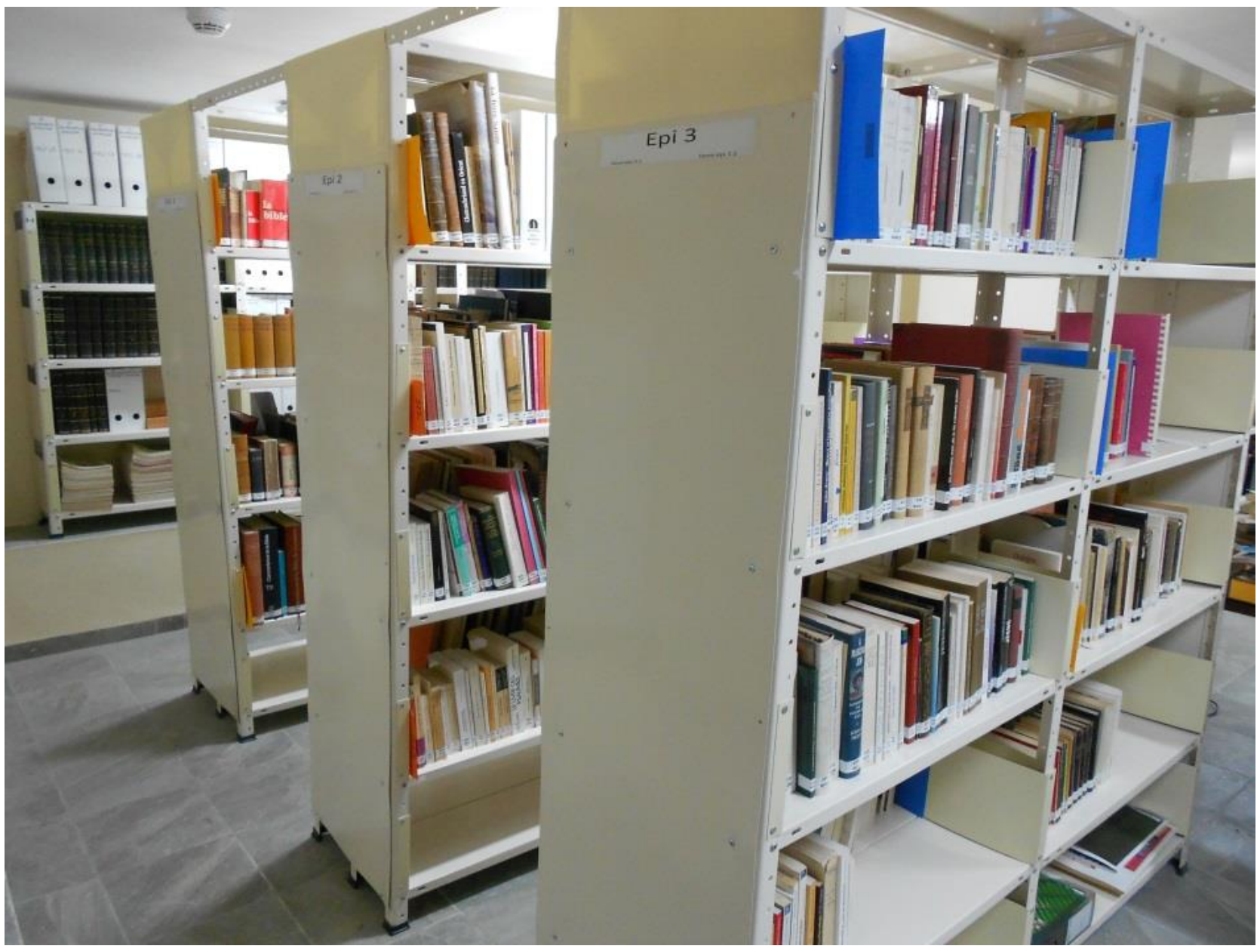

**Rapport technique**

**Jérusalem, août 2019**

**Rédigé par Benoit Soubeyran**

Bibliothécaire à l'Université de Paris-Est Marne-la-vallée, volontaire au sein de la Communauté assomptionniste de Saint-Pierre en Gallicante de juillet 2018 à janvier 2019 et de juillet à août 2019.

# Table des matières

Abréviations

BNF Bibliothèque Nationale de France

EBAF École biblique et archéologique française de Jérusalem

SPG Saint-Pierre en Gallicante

SUDOC Système Universitaire de Documentation

VIAF *Virtual International Authority File*

Remerciements

Je remercie les pères Jean Daniel Gullung et Cezar Andrei pour l'accueil et l'accompagnement des missions que j'ai réalisées à Saint-Pierre en Gallicante, ainsi que Marie-Claire Rouquet pour la relecture du présent travail. Les religieux, les salariés, les bénévoles et les hôtes avec qui j'ai eu la chance d'échanger à Saint-Pierre ont également enrichi et agrémenté positivement mon séjour à Jérusalem. Je remercie également le département des métadonnées de la BNF pour les corrections apportées à leur catalogue suite aux suggestions que j'ai formulées grâce au VIAF et à Wikidata.

# I. Présentation et historique

## Une bibliothèque chrétienne à Jérusalem

Les bibliothèques chrétiennes ont leurs origines dans la religion juive dont la pratique et la transmission dépendaient de la conservation et de la duplication des textes sacrés. Comme le judaïsme, le christianisme dépend fondamentalement de la préservation et de l'étude d'un texte sacré. Il s'ensuit que les textes et la littérature secondaire sont collectés à l'usage des communautés religieuses et transmis aux générations futures.

À Jérusalem, les bibliothèques religieuses ont une histoire ancienne. L'évêque Alexandre de Jérusalem a créé une bibliothèque pendant son mandat, dans la première moitié du IIIe siècle. C'est ce qu'indiquent les archives d'Eusèbe de Césarée, qui mentionne certaines des œuvres qu'il y a découvertes.

La bibliothèque de Saint-Pierre en Gallicante a connu une histoire mouvementée et différentes phases de classement. Elle s'est constituée par l'apport de différentes bibliothèques privées de religieux de Terre Sainte. Son histoire est intimement liée à la présence assomptionniste en Terre Sainte puisqu'une partie importante du fonds se trouvait dans le bâtiment Notre-Dame à Jérusalem (aujourd'hui *Notre Dame Center* tenu par les Légionnaires du Christ).

Trois noms de donateurs qui ont fait partie de la communauté assomptionniste de Saint-Pierre en Gallicante sont à signaler car leur nom revient souvent, marqué au crayon gris sur les premières pages des livres : celui du père Stanley (économe à l'époque où le père Adrien Masson était le supérieur), celui du père André Madec et surtout celui du père Alain Marchadour, exégète bibliste, dont la bibliothèque privée a connu une mésaventure. Ayant expédié par colis sa bibliothèque personnelle à la Communauté assomptionniste de Toulouse vers laquelle il avait été envoyé, elle est finalement revenue à Jérusalem, l'économe de la communauté de Toulouse n'étant pas au courant de l'arrivée du père Alain Marchadour. Elle est finalement restée à Saint-Pierre en Gallicante. Par la suite, les ouvrages de la bibliothèque ont longtemps dormi dans des cartons ou ont connu des déplacements hasardeux.

Deux séries de cotation ont été appliquées aux ouvrages : d'abord un classement alphanumérique de type A-1-1, A-1-2, … puis aujourd'hui un classement Dewey, avec indice numérique correspondant à une série thématique suivi des 3 premières lettres de l'auteur (ex. :

100.BERG pour un ouvrage de philosophie d'Henri Bergson). Les documents conservés sur l'ordinateur de la bibliothèque montrent plusieurs phases de classement depuis 2016.

## Les différentes phases de classement

### Mars et avril 2016 : premier modus operandi d'Anne Chevillard et de François Béjui

Tous les documents liés aux Augustins de l'Assomption auparavant cotés 255.4 (Code Dewey correspondant aux Augustins) ont vraisemblablement été re-cotés en AA, qui correspond aujourd'hui au fonds assomptionniste de SPG. Il a été établi de mettre en 921 – le code Dewey ne correspondant pas à cette thématique – les livres liés au patrimoine religieux, à la géographie, à la géopolitique de Jérusalem. Mais certains liés à la Terre Sainte ont été mis en 270 bis sans plus d'indication sur la distinction entre les deux séries. Les ouvrages sont encore listés sur des tableurs Excel. Les premières étiquettes et les premiers panneaux de signalétique ont été rédigés avec la répartition des cotes sur les épis. Aujourd'hui compte-tenu du désherbage effectué et du gain conséquent – et absolument nécessaire – d'espace occasionné, cette signalétique va devenir obsolète et devra être mise à jour.

### Avril 2017 : deuxième modus operandi de François Béjui

Le père Jean Daniel et François Béjui ont visité la Bibliothèque des Dominicains de l'EBAF le 6 avril 2017 pour s'inspirer de leur nomenclature et de leur organisation (un compte-rendu se trouve dans les documents). Mais pour la bibliothèque de SPG, s'inspirer de l'EBAF a ses limites : la bibliothèque du Couvent Saint-Étienne n'utilise pas un classement Dewey mais un classement « maison » utilisé nulle part ailleurs, le même dominicain cote les ouvrages depuis plus de douze ans. Le travail précédent de François Béjui a été continué sans changement majeur. C'est en mai 2017 qu'apparaît la première base de données Book'In avec 92 livres.

### Mars et avril 2018 : fiche explicative de Jacqueline Barro

Est ici repris une partie du document intitulé « fiche explicative » rédigé par Jacqueline Barro en avril 2018. J'y ai apporté quelques précisions pour la clarté de l'exposé. C'est en mars 2018 au cours d'une visite des Assomptionnistes à l'École biblique que j'ai pu avoir un premier contact avec Saint-Pierre en Gallicante.

#### *État des lieux au 1er mars 2018*

Les livres de la bibliothèque de Saint Pierre en Gallicante (SPG) ont fait l'objet de recensements (non exhaustifs) sur Excel. Au 1er mars 2018, les enregistrements réalisés portaient sur 2609 documents parmi lesquels 442 étaient classifiés « livres anciens », 171 « littérature », 1 906 « doc et essais » et 90 « périodiques ». SPG s'est ensuite doté d'une base de données, dite « Book'in », conçue spécialement pour la gestion de bibliothèques privées, et permettant notamment de répertorier, classifier, coder un fonds documentaire dans le double but d'aider les utilisateurs à rechercher un ouvrage et à le localiser en rayonnage.

Au 1er mars 2018, seuls les 442 livres anciens avaient fait l'objet d'une saisie dans ce nouveau logiciel et d'un rangement en bibliothèque.

### *D'Excel à Book'in*

Afin d'éviter une ressaisie du travail déjà réalisé sur les autres livres, l'option du transfert des données d'Excel vers Book'in a été privilégiée. Excel n'étant cependant pas directement compatible avec Book'in, il a été nécessaire de convertir en csv les trois fichiers concernés (doc et essais, littérature, périodiques).

Les transferts ont été difficiles, la structure des tableaux initiaux n'étant pas adaptée aux exigences de Book'In. Il en a résulté un long travail de vérification de presque toutes les informations transférées : noms des auteurs, titres des livres, maisons d'édition, lieux, date, etc. Travail fastidieux mais de nature à acquérir une assurance relativement raisonnable du bon transfert des données excel/csv.

### *Les constats du transfert*

Alors que 2 applications Book'in ont été téléchargées avec la volonté de continuer à distinguer les livres anciens des autres livres du fonds documentaire SPG, toutes les données ont fusionné dans une seule et unique base au moment du transfert. Une distinction est cependant toujours possible avec le champ « localisation » de Book'in[1]. Une requête permet d'isoler chacune des grandes catégories initialement prévues dans Excel (livres anciens, littérature, doc et essais, périodiques).

Le transfert a été fait sur la base des informations saisies dans Excel. Or le code associé à chaque livre n'est pas toujours, et est même rarement, celui retenu par les Pères pour le classement de leur bibliothèque. Le recensement Excel n'ayant pas couvert l'intégralité des documents de la bibliothèque, il reste à saisir dans Book'in tous les ouvrages non encore inventoriés.

### *La poursuite du travail*

Les Pères de SPG ayant opté pour un classement sur la base du code Dewey, il a paru normal de commencer par les livres dont ils avaient d'ores et déjà validé le code (livres figurant dans les étagères correspondant aux épis 1 à 7). A partir de l'épi 1, et jusqu'à l'épi 3, les livres ont été pris un à un afin de vérifier leur enregistrement dans Book'in, puis :

- pour les livres déjà enregistrés : le code de chacun a été vérifié et modifié quand il n'était pas conforme au classement retenu ;
- pour les livres non enregistrés : une saisie a été effectuée portant essentiellement sur le ou les noms d'auteur, le titre du livre, la maison, le lieu et la date d'édition, le code et son libellé ;
- les livres ont ensuite été classés par ordre alphabétique à l'intérieur de chaque code et sous code, puis rangés sur étagères.

Début avril 2018, la base Book'in contient 3365 documents (442 livres anciens, 2655 doc. & essais, 171 titres de littérature et 97 périodiques). Le travail de contrôle est à poursuivre à partir du code 233. Une réserve est toutefois portée sur l'ensemble du code 232 dont certains titres ne semblent pas correspondre au sujet du code.

[1] Le fonds de la Bibliothèque et celui des Livres anciens sont désormais sur deux bases séparées, ce qui respecte la volonté des Pères, le fonds des Livres anciens étant plus confidentiel que celui de la Bibliothèque. La séparation a été rendue nécessaire par les différences entre les deux fonds.

## *Le logiciel Book'In*

Deux icônes Book'in sont désormais sur l'écran d'accueil. L'un correspond à la Bibliothèque courante et l'autre aux Livres anciens. **Les deux bases ne peuvent être consultées en même temps et il faut veiller à fermer la base courante pour consulter l'autre.**

Avant d'accéder à la base des livres, le logiciel envoie un message pour proposer le rétablissement de liens invalides, cliquer sur « non ».

## *Après ouverture de la base de livres*

- Pour ajouter un livre, cliquer sur *livre* (en haut à gauche au-dessus de l'icône impression), puis sur *ajouter*. Une fenêtre s'ouvre pour la saisie des éléments du nouveau livre (renseigner les principaux champs Book'in, notamment ceux relatifs à l'auteur - si plusieurs auteurs, cliquer à droite et sur + à chaque nouvel auteur - au titre, aux conditions d'édition, et à la localisation (essais, littérature, périodique) ;
- Pour modifier un enregistrement, rechercher l'enregistrement à l'aide du moteur de recherche (loupe devant le rectangle blanc), sur-briller la ligne à modifier, cliquer droit pour modifier ;
- Pour dupliquer un livre, même procédure que pour modifier un enregistrement, et clic droit pour dupliquer et modifier les éléments non communs ;
- Pour rechercher un livre ou un ensemble de livres, cliquer sur la loupe, donner les critères de recherches (auteur, sujet d'un ouvrage, code, libellé, mots clés….). Indiquer les éléments sur lesquels la réponse devra porter (par exemple : auteur, titre, code, genre) ;
- Pour modifier le libellé d'un code, faire une recherche à partir du code, sélectionner tous les éléments de la réponse, et modifier le champ « genre » ;
- Pour enlever un livre, rechercher le livre par requête, surbriller, et enlever ;
- Pour sélectionner les éléments de réponse, cliquer sur la toute petite icône en haut à droite (sous la flèche), puis sur colonnes, puis sur les éléments à choisir.

Les tableaux CSV de l'import sont C:\Users\Bibliothèque\Desktop\Bibliothèque 2016-2018\2018-03 Bibliothèque Jacqueline. L'ensemble est sauvegardé sur le disque externe vert marqué de l'étiquette « Bibliothèque + Photos NDF ».

## Livres anciens : le travail de Nathalie

La base de données des Livres anciens est accessible par l'ordinateur de la bibliothèque. Une chose importante à signaler est que, pour consulter les notices avec les images reliées, il faut brancher le disque dur externe vert puisque les images sont liées aux notices par des adresses qui mènent au disque dur externe. **Le dossier intitulé « 602 PHOTOS Inventaire bibliothèque » du disque dur externe ne doit donc surtout pas être déplacé ou renommé.**

La particularité des Livres anciens est qu'ils se trouvent désormais sur une base de données séparée de la bibliothèque afin d'éviter les chevauchements des numéros d'inventaire. Cette séparation opérée en août 2019 explique pourquoi la numérotation des ouvrages de la bibliothèque comprend des zones blanches, entre le 0 et le 94, puis entre le 123 et le 443. Les numéros manquants sont à attribuer aux futurs livres à cataloguer. L'ordre des numéros a peu d'importance, car ils n'impactent pas la cote et le classement des ouvrages contrairement aux Livres anciens.

## De 2018 à 2019, le présent manuel de Benoit Soubeyran

Le travail de classification et d'informatisation que j'ai mené à la bibliothèque de Saint-Pierre en Gallicante s'est déroulé en trois phases :

- en mars 2018, j'ai aidé ponctuellement la bénévole Jacqueline Barro à paramétrer le logiciel Book'In et à transformer les tableurs Excel en fichiers .csv afin de faciliter l'import des notices dans le logiciel ;
- de juin à octobre 2018, j'ai assuré le catalogage et l'équipement des livres d'abord sous la direction du père Emmanuel Kahindo puis en autonomie, un problème de visa m'ayant contraint à rentrer en France en octobre 2018 et à travailler à distance pour le site web de Saint-Pierre en Gallicante ;
- de fin juillet à la mi-août 2019, j'ai pu achever une partie des tâches précédemment commencées - notamment le désherbage, avec le transfert des usuels les plus anciens et les plus volumineux vers le magasin – et le nettoyage de la base documentaire avec l'harmonisation des genres et des mots-clés.

Au 9 août 2019, la bibliothèque de Saint-Pierre en Gallicante, compte 3388 ouvrages dans le catalogue. Plus de la moitié possède une cote qui permet de les localiser avec exactitude. Compte-tenu du nombre d'ouvrages qui ne possède pas de notice, le fonds de la bibliothèque peut être estimé à plus de 3500 ouvrages. En plus des livres de la Bibliothèque, il faut rajouter les 622 ouvrages du fonds des Livres anciens catalogués par Nathalie.

Les tâches que j'ai menées ont constitué un travail de longue haleine qui a souvent nécessité de faire une pause dans le catalogage répétitif des livres. Très souvent, il m'est apparu nécessaire de bien réfléchir à la taxinomie. En effet un choix judicieux de mots-clés a été et est la condition *sine qua none* d'une bonne classification. Malgré les difficultés rencontrées, le plaisir de voir avancer le classement de cette bibliothèque fut pour moi un plaisir et une vraie source de joie. Je peux tout à fait faire mienne cette citation d'Hermann Hesse :

> *C'est un bonheur que de se constituer progressivement une belle petite bibliothèque avec des moyens modestes en bravant toutes les difficultés ; c'est un sport passionnant !*[2]

## Le volontariat à Jérusalem

Faire un volontariat à l'étranger est toujours une expérience unique, riche de rencontres, de découvertes et d'enseignement. Le faire ici à Jérusalem est d'autant plus enrichissant que la Terre Sainte et un territoire aux racines anciennes et à l'histoire complexe. La profusion des lieux de culte et des sites historiques y est immense. Jérusalem agit en effet comme un centre d'attraction pour

[2] Hermann Hesse, *Une bibliothèque idéale*, éd. Payot & Rivages, 2012 [lire en ligne sur Google Books] (consulté le 9 août 2019).

l'univers chrétien, où l'Orient et l'Occident s'y rencontrent dans un baiser fraternel.

Parmi les souvenirs de mon volontariat, je me rappelle avoir été invité le samedi 29 septembre 2018 par les deux étudiants assomptionnistes de Saint-Pierre en Gallicante, j'ai pu assisté à la messe de rentrée académique du *Studium Theologicum Salesianum* du monastère Ratisbonne[3]. Fondé au siècle dernier par les deux frères Ratisbonne, c'est aujourd'hui un campus cosmopolite avec des étudiants dont la vocation est de servir l'Église. Ils viennent des quatre coins du Monde, du Vietnam à l'Afrique en passant par l'Italie et l'Amérique du Sud.

Cet événement fut aussi l'occasion de visiter la riche bibliothèque universitaire. Moi-même bibliothécaire au sein de la communauté assomptionniste de Saint-Pierre en Gallicante, je ne peux rester insensible au fait de retrouver ici à Jérusalem tant de lieux de prière associés à des lieux de savoir : la méditation est une activité indissociable de celle de l'étude.

## Le rôle des bibliothèques chrétiennes

Le rôle et la fonction des bibliothèques chrétiennes ont toujours été caractérisés par la continuité dans l'agitation, l'ingéniosité dans des ressources souvent insuffisantes, et la promotion de ce qui a une valeur durable dans un contexte de changement ecclésial et sociétal constant. Selon les mots de Cassiodore :

> *Nous visons à la fois à préserver ce qui est ancien et à construire quelque chose de nouveau ; nous désirons élever des choses qui sont modernes sans diminuer les œuvres de nos ancêtres*[4].

Ce n'est qu'en de très rares occasions que l'héritage et les contributions des bibliothèques chrétiennes ont été remarqués, et généralement longtemps après coup. L'ouvrage de Thomas Cahill *How the Irish Saved Civilization* fournit néanmoins une agréable exception. Depuis Eusèbe de Césarée, les bibliothèques théologiques ont le plus souvent été mises en valeur par les efforts de ceux qui savourent un certain degré d'anonymat.

L' « ère de l'information » est communément considérée comme une menace grave à la perpétuité des textes imprimés et des collections sur papier des bibliothèques. Mais, bien sûr, elle promet aussi une collaboration sans précédent entre les excellentes collections et les bibliothécaires astucieux qui y travaillent. Il y a donc lieu d'espérer que les meilleurs jours pour les bibliothèques chrétiennes ne se situent pas dans le passé mais dans l'avenir[5].

# II. Catalogage et indexation dans Book'In

Dans le cadre de travail de catalogage et d'indexation de la Bibliothèque de Saint-Pierre en Gallicante, il a fallu créer de nouvelles notices d'auteur et de livre et corriger celles existantes. Pour ce travail, il est primordial de s'appuyer sur les informations contenues dans l'ouvrage.

---

[3] Benoit Soubeyran, « Le Studium Theologicum Salesianum et sa bibliothèque », sur benoitsoubeyran.wordpress.com, 2 octobre 2018 (consulté le 13 août 2019)r

[4] Cité dans Richard Southern, "A Benedictine Library in a Disordered World", in *Downside Review*; n° 94, Juilllet 1976, p. 169.

[5] David Stewart, « Christian Libraries », In *International Dictionnary of Libray Histories*, Fitzroy Dearborn, London, p. 48-54.

Mais il est tout aussi primordial - afin que le travail soit de qualité - de s'appuyer sur les catalogues et les bases de données en ligne. Quatre outils ont été d'une aide inestimable pour accompagner la réalisation de ma mission au sein de la Bibliothèque de SPG :

- Le VIAF, pour les notices d'autorité [https://viaf.org/] ;
- Wikidata, notamment pour les dates et lieux de naissance/décès des auteurs [https://www.wikidata.org] ;
- Le catalogue de la BNF, à la fois pour les notices d'ouvrage et d'autorité [https://catalogue.bnf.fr/index.do] ;
- Le SUDOC en complément, notamment lorsque la notice d'ouvrage était manquante sur la BNF [http://www.sudoc.abes.fr/].

Le travail de catalogage a ainsi permis de signaler au département des métadonnées de la BNF les doublons d'autorité et d'améliorer ainsi la qualité du catalogue par le dédoublonnage des notices[6].

Les informations tirées des ouvrages et des bases de données en ligne ont ainsi permis de remplir précisément les champs des notices du logiciel Book'In. La mission que j'ai accomplie au sein de la bibliothèque de SPG m'a permis de concilier mes compétences de bibliothécaire et de wikimédien. On peut espérer qu'à l'avenir les bibliothécaires et documentalistes auront acquis suffisamment d'expérience pour tirer le maximum d'informations des outils de la galaxie wikipédia, tout en étant capable d'apporter les corrections nécessaires sur ces outils.

## Base des auteurs

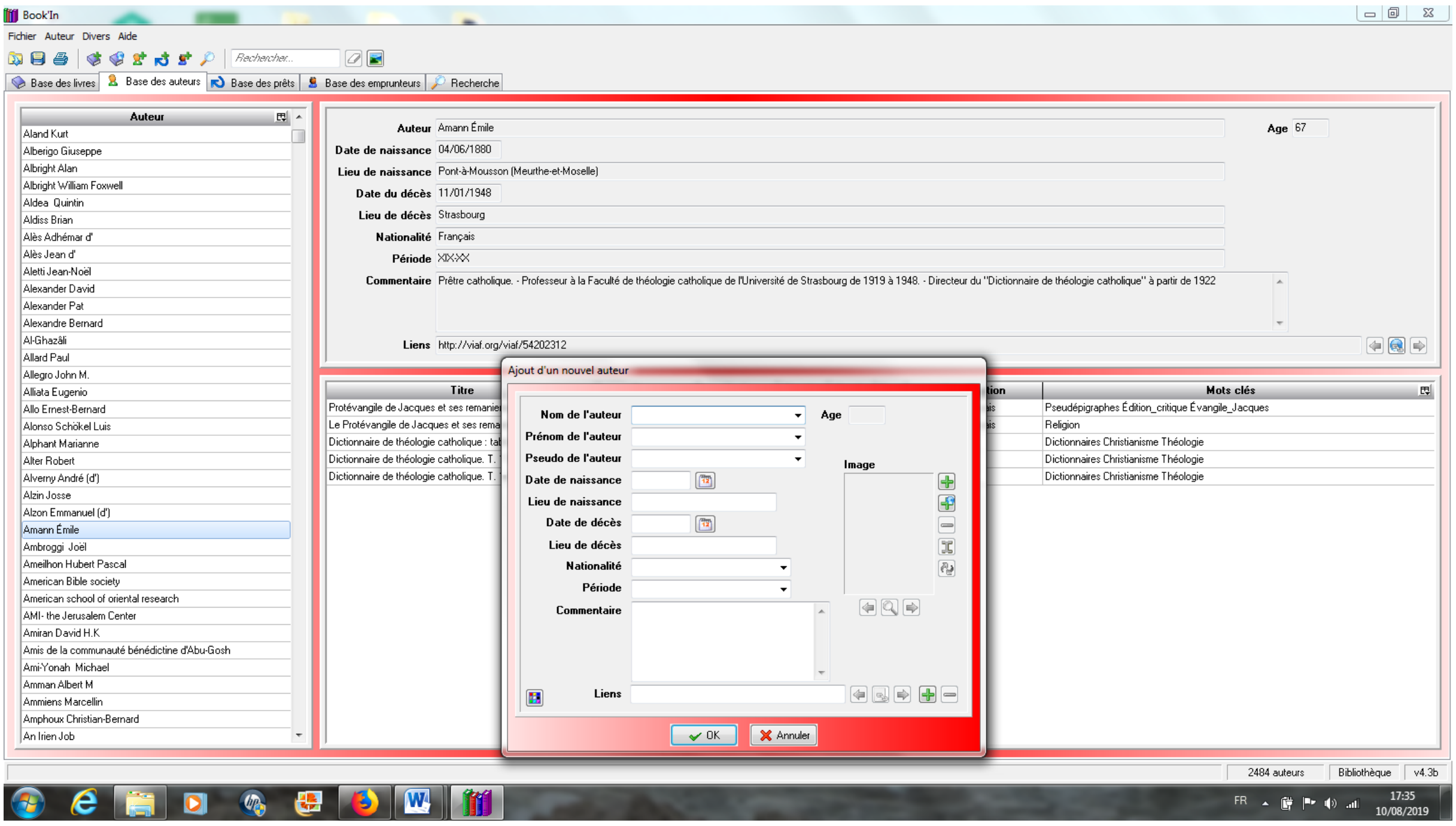

---

6 Benoit Soubeyran, « Le catalogage et wikidata aident à corriger les notices de la BNF », sur benoitsoubeyran.wordpress.com, 15 mars 2018 (consulté le 9 août 2019)

Pour compléter au mieux, les notices d'auteur, il faut aller chercher les informations contenues dans le VIAF, sur les notices d'autorité de la BNF et dans Wikidata.

### Nom de l'auteur

Première lettre en majuscule. Pour des noms avec des particules, mettre la particule dans le champ prénom. Exemple : Nom : Alès ; Prénom : Jean d'. Remplir aussi dans ce champ la collectivité. Exemple : Conférence des évêques de France.

### Prénom de l'auteur

Première lettre en majuscule

### Pseudo de l'auteur

Surtout dans le cas où le pseudonyme est plus connu et usité que le nom véritable, comme pour Daniel-Rops. Exemple : Nom : Wojtyla ; Prénom : Karol ; Pseudonyme : Jean-Paul II

### Date de naissance

Sous le format Jour/Mois/Année.

### Lieu de naissance

Commune de naissance. Précision de la localisation administrative si nécessaire : Département français, État des États-Unis, etc.

### Date de décès

Sous le format Jour/Mois/Année.

### Lieu de décès

Commune de décès. Précision de la localisation administrative si nécessaire : Département français, État des États-Unis, etc.

### Nationalité

Français, Allemand, etc.

### Période

En fonction de la date de naissance et de décès. Prendre aussi en compte la période d'activité.

### Commentaire

Le plus souvent, copier/coller la note sur l'œuvre de la notice de la BNF.

### Liens

Mettre le lien VIAF correspondant.

### Age

Il est calculé automatiquement avec la date de naissance et la date de décès.

## Base des livres

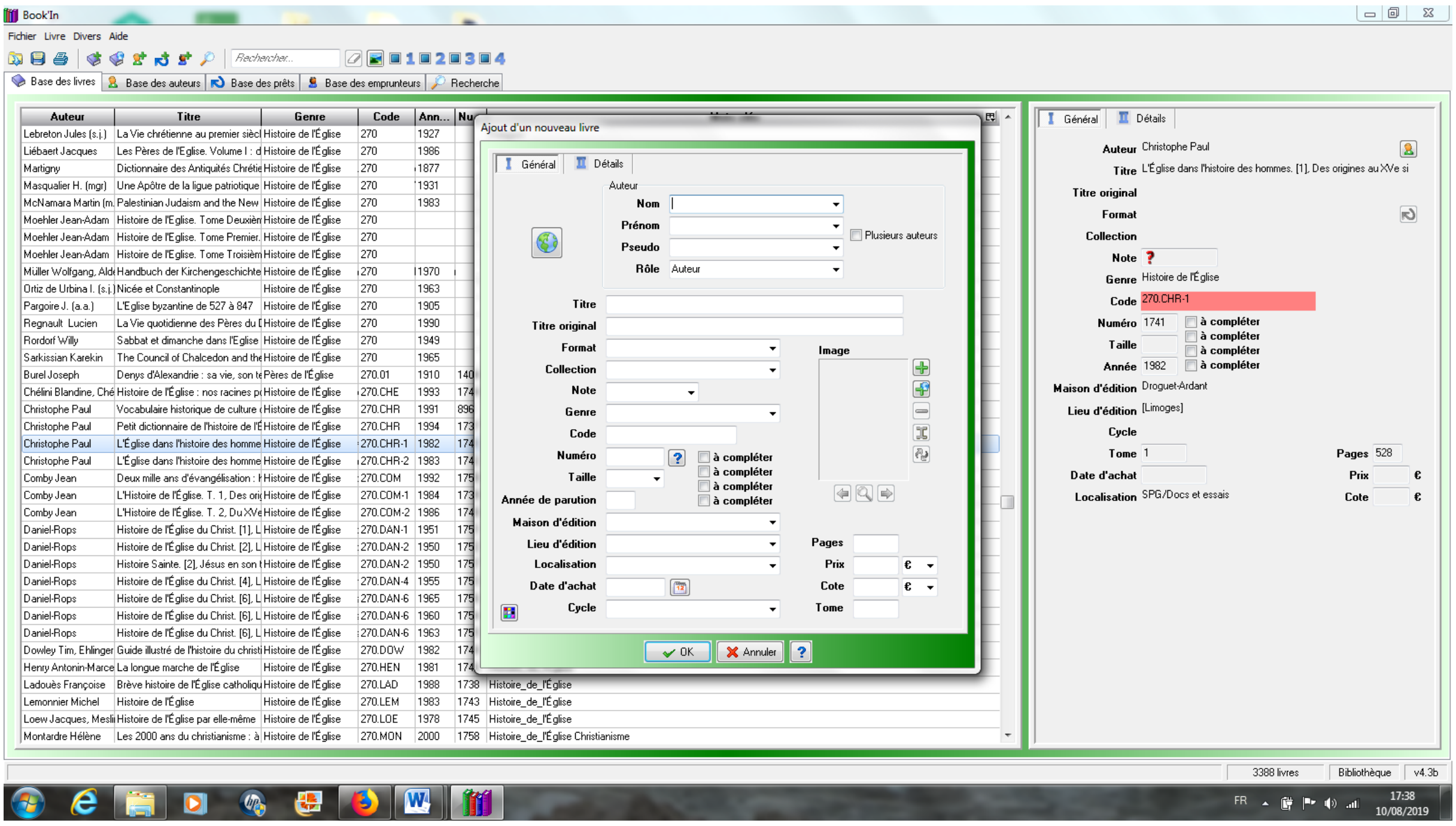

### I Général

#### *Champs auteur*

Commencer par le nom ou par le pseudonyme. Cocher « plusieurs auteurs ».

#### *Titre*

Première lettre en majuscule. Séparer le titre et le sous-titre par « : ». Pour les monographies en plusieurs tomes, voir l'exemple du livre de Jean Comby :

- L'Histoire de l'Église. T. 1, Des origines au XVe siècle ;
- L'Histoire de l'Église. T. 2, Du XVe au XXe siècle.

#### *Titre original*

Mettre le titre original si l'ouvrage est une traduction.

#### *Format*

Laisser le champ vide.

#### *Collection*

Première lettre en majuscule. Inutile de mettre le numéro.

#### *Note*

Laisser le champ vide.

#### *Genre*

Mettre le genre correspondant à la classe de la classification décimale de Dewey (Voir III Classement Dewey).

### *Code*
Les chiffres de la classe Dewey, suivi d'un point et des trois premières lettres du nom d'auteur. L'inscrire au crayon gris sur la première page de l'ouvrage catalogué. Commencer par la lettre M, si le livre doit être rangé dans le magasin.

### *Numéro*
Cliquer sur le point d'interrogation pour affecter un numéro d'inventaire à l'ouvrage et l'inscrire au crayon gris sur la première page entre crochets.

### *Taille*
Laisser le champ vide.

### *Année de publication*
Mettre l'année de publication et non d'impression.

### *Maison d'édition*
Première lettre en majuscule. Si plusieurs éditeurs, les séparer par une virgule.

### *Lieu d'édition*
Première lettre en majuscule.

### *Localisation*
Pour la plupart des monographies, choisir SPG/Docs et essais.

### *Date d'achat*
Laisser le champ vide.

### *Cycle*
Laisser le champ vide.

### *Prix*
Laisser le champ vide.

### *Cote*
Laisser le champ vide.

### *Tome*
Mettre le numéro du tome en chiffres indo-arabes, s'il s'agit d'une monographie en plusieurs volumes.

## II Détails

### *Commentaire*
Mettre des remarques pertinentes dans ce champ.

### *Liens*
Insérer le lien vers la notice correspondante.

### Dimensions

Mettre la hauteur du livre en millimètre.

### Poids

Laisser le champ vide.

### Résumé

Laisser le champ vide.

### Mots clés

Mettre les mots-clés séparés par un espace. Faire attention, car dans le cas d'un mot-clé entrecoupé d'espaces, utiliser le tiret bas ou *underscore* : « Histoire_de_l'Église » et non « Histoire de l'Église », car sinon, le logiciel Book'In enregistre trois mots-clés « Histoire » « de » et « l'Église ».

## III. Classement Dewey

La classification décimale Dewey (CDD) est structurée autour de dix classes principales couvrant l'ensemble du monde du savoir. Chaque classe principale est ensuite structurée en dix divisions hiérarchiques, chacune ayant dix sections de spécificité croissante. En tant que système de classification des bibliothèques, la CDD classe par discipline et non par sujet, de sorte qu'un sujet comme les vêtements est classé en fonction de son traitement disciplinaire (influence psychologique des vêtements à 155,95, coutumes associées aux vêtements à 391, et mode des vêtements à 746,92) dans le cadre du concept.

Pour le classement de la bibliothèque de Saint-Pierre en Gallicante, ce sont surtout les classes et les sous-classes liées à la religion (classe Dewey 200) qui ont été utilisées. Le livre d'Annie Béthery, *Abrégé de la classification décimale de Dewey* dans son édition de 1990, a constitué plus qu'un simple ouvrage de consultation, il a servi d'instrument de travail tout au long du catalogage[7]. Par ailleurs le fichier pdf consultable sur le site *Regnat* a été un appoint appréciable lorsque les sous-classes étaient insuffisamment détaillées dans l'ouvrage d'Annie Béthery[8].

Le choix de mots-clefs apposés aux ouvrages va être déterminant dans le catalogage puisque c'est le premier mot qui va déterminer le choix du genre et de la classe Dewey correspondante. Il faut arriver presque au schéma d'une équation mathématique pour obtenir la classification optimale.

### 020 Sciences de l'information

Sciences_de_l'information

et/ou

---

[7] Annie Béthery, *Abrégé de la classification décimale de Dewey : nouvelle édition augmentée à partir de la première version intégrale française et de la XIXe édition intégrale en langue anglaise*, éd. Du Cercle de la Librairie, Paris, 1990, 263 p. [cote 020.BET et n° d'inventaire 1761 dans la Bibliothèque de Saint-Pierre en Gallicante].

[8] https://regnat.pagesperso-orange.fr/Doc000/Dewey.pdf (consulté le 12 août 2019).

Bibliothèque

## 030 Encyclopédies

Encyclopédies

## 100 Philosophie

Philosophie

## 203 Dictionnaires du christianisme

Dictionnaires + Christianisme

## 220 Bible

Bible

### *220.3 Dictionnaires de la Bible*

Bible

+

Dictionnaires ou Encyclopédies

### *220.41 Versions originales de la Bible*

### *220.42 Manuscrits de la mer Morte*

Qumran

### *220.5 Versions modernes de la Bible*

### *220.6 Critique de la Bible*

### *220.70 Commentaires de la Bible (introductions)*

Le père Emmanuel Kahindo a voulu opérer cette distinction entre les introductions à la Bible (220.70) et les Commentaires (220.71). Avec le recul, cette distinction n'apparaît pas forcément judicieuse car elle est très difficile à faire apparaître au niveau de l'indexation.

### *220.71 Commentaires de la Bible*

### *220.9 Bible et Histoire*

#### 220.91 Atlas, albums, guides

#### 220.92 Archéologie biblique

Archéologie

+

Bible (Ancien/Nouveau testament)

et/ou

Terre_Sainte

220.93 Bible et Voyages

Voyages + Bible

## 222 Livres historiques de l'Ancien Testament

## 223 Livres poétiques de l'Ancien Testament

## 224 Livres prophétiques de l'Ancien Testament

## 226 Évangiles et Actes des Apôtres

Évangiles

### *226.1 Évangile de Matthieu*

Évangile_Mathieu

### *226.2 Évangile de Marc*

Évangile_Marc

### *226.3 Évangile de Luc*

Évangile_Luc

### *226.4 Évangile de Jean*

Évangile_Jean

### *226.5 Actes des Apôtres*

Actes_des_Apôtres

## 227 Épîtres (incluant les Épîtres de Paul)

Épîtres

ou

Simon-Pierre

ou

Épîtres_Paul

## 228 Apocalypse

Apocalypse

## 229 Livres apocryphes

Pseudépigraphes

# 230 Théologie chrétienne

Théologie_chrétienne

## 231 Dieu

Dieu

## 232 Jésus-Christ

Jésus-Christ

### *232.91 Marie*

Marie

### *232.97 Passion et Résurrection*

Passion

Ou

Résurrection

Ou

Saint_Suaire

## 235 Hagiographie

Hagiographie

ou

Bienheureux + Biographies

## 241 Théologie morale

Théologie_morale

## 242 Prière et dévotion

Prière

## 252 Sermons

Sermons

## 261.2 Dialogue_interreligieux

Dialogue_interreligieux

## 261.55 Religion et sciences

Religion + Sciences

## 262 Ecclésiologie

Ecclésiologie

### *262.001 Œcuménisme*

Œcuménisme

### *262.13 Papes*

Papes

### *262.8 Conciles*

Conciles

Ou

Vatican_II

### *262.9 Droit canonique*

Droit_canonique

### 264 Liturgie

Liturgie

### *264.01 Liturgie orientale*

Liturgie_orientale

### *264.02 Liturgie catholique*

Liturgie_catholique

### *264.3 Eucharistie*

Eucharistie

### 265 Sacrements

Sacrements

### *265.1 Baptême*

Baptême

### *265.5 Mariage*

Mariage

### *265.6 Pénitence*

Pénitence

### *265.7 Onction des malades*

Onction_des_malades

### 268 Catéchèse

Catéchèse

## 270 Histoire de l'Église

Histoire_de_l'Église

### 270.1 Église primitive (30-600)

Église_primitive

### *270.11 Pères de l'Église*

Pères_de_l'Église

### 270.2 Église médiévale (600-1500)

Église_médiévale

### 270.3 Église moderne (1500-...)

Église_moderne

## IV. Index des mots-clés

## V. Wikidata et le VIAF : des outils pour dédoublonner les notices d'autorité de la BNF

| | | | | | Notices BNF dédoublonnées | | |
|---|---|---|---|---|---|---|---|
| | **Date** | **Auteur** | **VIAF** | **WIKIDATA** | **id BNF 1** | **FRBNF 2** | |
| | 2018 | | | | | | |
| 1 | 8/3 | Ferdinand Brettes | 24592421 | Q49767777 | 16477403z | FRBNF10649547 | |
| 2 | 15/3 | Hésychius d'Alexandrie | | Q443510 | 2682950n | FRBNF12682950n | |
| 3 | 3/5 | Jézabel | | Q721295 | 135560258 | FRBNF17077899c | |
| 4 | 2/7 | Ernst Haenchen | 110249955 | | 12685885r | FRBNF11487486 | |
| 5 | 16/7 | Evode Beaucamp | | | 13526371n | FRBNF118907621 | |
| 6 | 19/7 | William D. Davies | 61567153 | | 10849470t | FRBNF1205932 | |
| 7 | 27/7 | Cyrus I. Scofield | 20482704 | Q768465 | 116345369 | FRBNF10536648 | |
| 8 | 28/7 | Joseph Verdunoy | | | 116349383 | FRBNF12993148 | |
| 9 | 7/8 | Léon Cré | 29722573 | | 10390941m | FRBNF11054702 | |
| 10 | 7/8 | Henri Nicole | 15074128 | | 15342135m | | |
| 11 | 9/8 | Pierre Duvignau | 88952234 | | 11579616t | FRBNF11338937 | |
| 12 | 20/8 | Hilda F. M. Prescott | 116548338 | Q187330 | 11319540t | FRBNF11082097 | |
| 13 | 21/8 | Charles Warren | 40318203 | Q74338 | 106121454 | FRBNF15343453 | |
| 14 | 21/8 | Xavier Marchet | 317193203 | | 10675967m | FRBNF11269639 | |
| 15 | 21/8 | J. T. de Belloc | 26819430 | | 127370646 | FRBNF10299904 | |
| 16 | 21/8 | David Roberts | 32031417 | | 16487109d | FRBNF12134627 | |
| 17 | 6/9 | Jules Grimal | 290764332 | | 11222968n | FRBNF11840357 | |
| 18 | 7/9 | Raymond Thibaut | 29676803 | | 13323569k | FRBNF13014811 | |
| 19 | 7/9 | Joseph Gill | 112990216 | Q47490651 | 115564958 | FRBNF11254579 | |
| 20 | 7/9 | Jean-Baptiste Chautard | 29531782 | | 11332390j | FRBNF11896401 | |
| 21 | 13/9 | Claude Cuénot | 109757666 | | 118982690 | FRBNF15457318 | |
| 22 | 14/9 | George Périès | 56635206 | Q28843391 | 120830433 | FRBNF11082410 | |
| 23 | 14/9 | Gervais Dumeige | 56609071 | | 11901086n | FRBNF11567095 | |
| 24 | 21/9 | Michel Breydy | 93907883 | | 120534307 | FRBNF10938608 | |
| 25 | 21/9 | Michel Hayek | 8572684 | | 121759749 | FRBNF10903261 | |
| 26 | 21/9 | Peter Hebblethwaite | 110596119 | Q2048220 | 12612357z | FRBNF11572405 | |
| 27 | 25/9 | Franck Edward Brightman | 37173690 | | 112638929 | FRBNF14639628 | |
| 28 | 26/9 | Congregatio pro Ecclesiis… | 139486695 | Q577568 | 11742463t | FRBNF13569494 | |
| 29 | 28/9 | Alphonse Raes | 20823874 | Q46011944 | 115638110 | FRBNF11032194 | |
| 30 | 28/9 | Bernard D. Marliangeas | 4933292 | | 11914642m | FRBNF14119035 | |
| 31 | 28/9 | Emanuele Testa | 12351487 | Q9253374 | 121828437 | FRBNF11554700 | |
| 32 | 30/10 | Daniel Brugès | 8304 | | 154526163 | FRBNF11894261 | |
| 33 | 27/11 | David von Günzburg | 3592366 | Q5234499 | 10568626j | FRBNF12899948 | |
| | 2019 | | | | | | |
| 34 | 29/7 | Michel Lemonnier | 05233994 | Q65966925 | 126203659 | | |
| 35 | 30/7 | Jacques Montjuvin | 5859934 | | 12626684v | FRBNF10478969 | |
| | | | | | | | |

Le travail de catalogage à la bibliothèque de Saint-Pierre en Gallicante m'a permis de repérer et de signaler les doublons des notices d'autorité de la BNF. La liste suivante énumère les 35 notices d'autorité qui ont pu être corrigées par le département des métadonnées de la BNF depuis le 8 mars 2018.La procédure de dédoublonnage avait déjà été signalée dans un article de mon blog en mars 2018[9].

Wikidata permet en effet la centralisation des notices d'autorité. Il est facile pour un utilisateur confirmé de détecter les doublons et de fusionner des éléments. Pour éliminer davantage de doublons, il faut souhaiter que davantage d'auteurs soient présents sur Wikidata.

[9] Benoit Soubeyran, « Le catalogage et wikidata aident à corriger les notices de la BNF », sur benoitsoubeyran.wordpress.com, 15 mars 2018 (consulté le 23 août 2019).

L'auteur

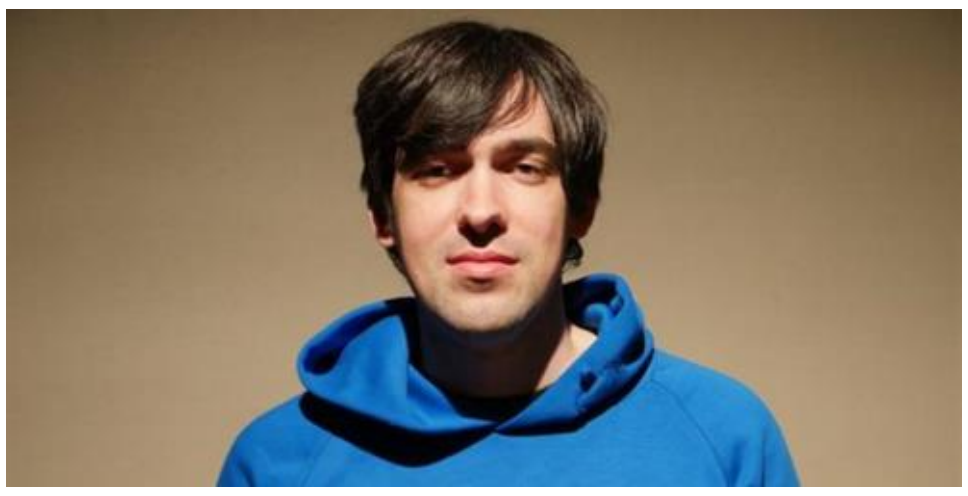

**Benoit Soubeyran** est né le 7 octobre 1986 à Nîmes. Au cours de ses études d'histoire et de lettres, il a travaillé comme bibliothécaire à l'Université Paul-Valéry de Montpellier de 2011 à 2012. Il a également été archiviste pour la Ville de Montpellier en 2014 et pour l'Agence d'urbanisme de Nîmes en 2015. Bibliothécaire volontaire à Jérusalem en 2018, il est depuis 2019, bibliothécaire à l'Université de Paris-Est Marne-la-Vallée (source : benoitsoubeyran.wordpress.com [consulté le 10 août 2019]).

Bibliographie de l'auteur